\documentclass[onecolumn,11pt]{article}
\def \degC {~^{\rm{o}}}	
	
\usepackage{graphics}
\usepackage[square,comma,longnamesfirst,numbers,sort&compress]{natbib}
\usepackage[pdftex]{graphicx}
\usepackage[in]{fullpage}

\usepackage{amssymb,amsfonts,amsmath}
\usepackage[nomarkers]{endfloat}

\begin{document}
	

	
\title{Electrical properties of polyetherimide thin films:\\ {\Large Nonparametric dielectric response analysis with distribution of relaxation times}}

\author{Enis Tuncer\\
	{\textit{General Electric Global Research, Niskayuna 12309 New York USA}}}
	
\maketitle
	
	    
\begin{abstract} 
High temperature polymeric materials for electrical insulation and energy storage are needed for transformational power applications such as pulsed-power and hybrid electrical vehicles. One of the candidate materials has been polyetherimide, an amorphous thermoplastic with a glass transition over 200$\degC$C. Here, the dielectric studies on the material are reported by taking into account the polarization and conduction processes in the polyetherimide. The dielectric data were analyzed with the distribution of relaxation times approach with Debye expression as the kernel in the inversion algorithm, the results then created the relaxation map for polyetherimide. The method satisfies the Kramers-Kronig relationships, so the ohmic conductivity and permittivity at high frequencies could be estimated from the experimental data even there exists significant measurement error. The data were compared to the published results on polyetherimide in the literature. The materials is a low loss polymer with negligible ohmic losses below  200$\degC$C. The estimated fragility of the polyetherimide was high about 284 from the resolved relaxation map using Vogel-Tammann-Fulcher expression. Polyetherimide is a good dielectric for advanced energy storage and electrical insulation technologies up to 200$\degC$C.

\begin{itemize}
\item[{\em Keywords}] Polymer physics, dielectric relaxation, polyetherimide, inverse problem, dielectric permittivity, thin films, fragility
\end{itemize}
\end{abstract}
	
\section{Introduction}
General Electric developed polyetherimide in its Corporate Research Center in early 1970's \cite{UltemPatent1,UltemPatent2}. The material is a high performance amorphous thermoplastic (class H insulation, 180$\degC$C operation temperature) dielectric with low losses, \cite{Shugg1986} and with high glass transition temperature $\sim215^o$C \cite{Belana1998Part1,Diaz1998Part2}. Different grades of polyetherimide have been derived for specific applications \cite{Krause1998}. Its low loss behavior and high glass transition temperature property make it a valuable alternative to some of the conventional electrical insulation materials and dielectrics, i.e., polypropylene based thin film electrostatic capacitors \cite{Sarjeant1998,Reed1994,Reed1992}. Polymeric thin film based electrostatic capacitors require materials with high dielectric permittivity, low dielectric losses (see Sarjeant et al. \cite{Sarjeant1998} and Reed and Cichanowski \cite{Reed1994} for reviews on capacitor technology). One should not forget the process-ability of the film and working range of the polymer;  and high temperature capability. Conventional polymeric film capacitors are mainly designed with polypropylene, polyethylene teraphthalate \cite{Reed1992} and polycarbonate films. 

Previous studies on polyetherimide has indicated that depend ending on the grade of the material \cite{Krause1998} and its moisture content \cite{Fontanella2007}, dielectric behavior as a function of temperature and frequency were different. 
Molecular mobility, chain dynamics and merging of the segmental local motion ($\beta$-relaxation) and cooperative process ($\alpha$-relaxation) are studied through the dynamic glass transition \cite{GlassTrans,LunkenheimerBDSBook} via dielectric or impedance spectroscopy \cite{KremerBook,MacdonaldImp2}. We determine the charge transport just before and after the glass transition. The behavior of charge mobility and polarization determines the electrostatic charging of the film and how it can be dissipated during the production of capacitor devices, \cite{Blaise1990,NielsElectrostatics}. Special precautions needs to be taken to abate static charge on near-to-perfect insulating dielectrics films.

Here we concentrate on the dielectric relaxation properties of an amorphous polymer, polyetherimide, for dielectric applications. The paper is organized as follows; a brief introduction to distribution of relaxation times approach is given in \S~\ref{sec:distr-relax-times}; The information on polyetherimide and the measurements are described in \S~\ref{sec:materials-methods}; Results on electrical properties and resolved distribution of relaxation times with information on the fragility are presented in \S~\ref{sec:electr-prop-poly}; Finally a discussion and conclusions are presented in \S~\ref{sec:disc-concl}.

\section{Distribution of relaxation times approach}\label{sec:distr-relax-times}	

\subsection{Preliminary results}
In order to separate the frequency independent intrinsic properties from those of frequency dependent, ones a method based on the distribution of relaxation times was adopted \cite{Frohlich,Yager1936,Kauzmann,DRTMacdonals1963,Jonscher1983,McCrumDRT}. Different approaches have been suggested to resolve the distribution function from experimental data in the literature \cite{Kliem88,Grimau1999,Keiter98,Dias,HNdist,Arndt1,Macdonald2000a,Macdonald2000b,macdonald:6241}. The method developed to resolve the relaxation times by the present author is a non-parametric approach \cite{Tuncer2000b,TuncerJAP2006,TuncerMacdonald2007}, which does not suffer from initial guesses, {\em a-priori} assumptions, such as selection of empirical expression(s) or model and predictions for the model parameters. The method resolves continues distributions and applied to many different systems previously (see the literature \cite{TuncerINTECH2011,TuncerMacdonald2007,Nanotech2006,TuncerPhilMag2006,TuncerJAP2006,TuncerJNCS,TuncerJCP2005,Tuncer2004a,Tuncer2002b,Tuncer2002elec,TuncerPhD,Tuncer2000b}). 

\subsection{Dielectric  representation} 

The relative complex dielectric permittivity of materials $\varepsilon^*$ as a function of state variables (angular frequency $\omega$, temperature $T$, pressure $P$, etc.) should be expressed as follow,
\begin{eqnarray}
  \label{eq:eps}
  \varepsilon^*(\omega,T,P,\dots)=\chi^{*}(\omega,T,P,\dots)+\varepsilon_{\infty}(T,P,\dots)+\sigma_0(T,P,\dots)(\imath\varepsilon_0\omega),
\end{eqnarray}
here, $\chi^*$ is the complex dielectric susceptibility; $\varepsilon_{\infty}$ is permittivity at optical frequencies or higher frequencies than the probed frequency region; $\varepsilon_0$ is the permittivity of vacuum, $\varepsilon_0=8.854$pFm$^{-1}$; and $\sigma_0$ is the ohmic conductivity of the material. For low loss dielectric or good electrical insulators to resolve their low conductivity one needs to perform measurements at low frequencies where the ohmic loss contribution becomes significant compare to the polarization losses. 

In the adopted formalism in the data analysis method, the complex permittivity is expressed as,
\begin{eqnarray}
   \varepsilon^*(\omega,T)&=&\chi^{*}(\omega,T)+\varepsilon_{\infty}(T)+\sigma_0(T)(\imath\varepsilon_0\omega)^{-1},\quad \text{with} \label{eq:chi}\\
	\chi^*(\omega,T)&=&\int_0^\infty {\sf g}(\tau,T)(1+\imath\omega\tau(T))\rm{d}\tau,\label{eq:drt}
\end{eqnarray}
where ${\sf g}(\tau)$ is the distribution of relaxation times and $\tau$ is the time constant. The frequency dependent term on the right-hand-side inside the parenthesis in Eq.~(\ref{eq:drt}) is the kernel, and it is the response of a single dipole derived by \citet{Debye1945} (see also \citet{MacdonaldDebye} and \citet{Frohlich}. Observe that the two material parameters $\varepsilon_{\infty}(T,P)$ and $\sigma_0(T,P)$ are frequency dependent. The numerical procedure yields the distribution of relaxation times ${\rm g}(\tau)$, the high frequency permittivity $\varepsilon_\infty$ and the ohmic conductivity $\sigma_0$. Solution and numerical details to this inverse problem and other  problems in dielectrics are discussed elsewhere \cite{Tuncer2001d,TuncerSpectralPRB,TuncerJAP2006,TuncerLang,TuncerLang2008,TuncerTDEI2012,TuncerISE2011,TuncerMacdonald2007}.

The most probable relaxation time $\bar{\tau}$ for a process is estimated with 
\begin{eqnarray}
  \label{eq:mostexptau}
  \bar{\tau}=\left[{\int_a^b \tau\times{\sf g}(\tau){\rm d}\tau}\right]\times\left[{\int_a^b{\sf g}(\tau){\rm d}\tau}\right]^{-1}
\end{eqnarray}
The integral limits $(a,b)$ were taken from the corresponding time axis ($\sim\omega^{-1}$) for the measurement window frequencies.

\section{Materials and methods}\label{sec:materials-methods}

 The studied film was a 5$\mu$m polyetherimide film (commercially known as Ultem\textregistered 1000 from SABIC Innovative Plastics\texttrademark).  The studied film was supplied from Shin-Etsu Polymer Co., Ltd Japan.

We have employed a Novocontrol Quatro Broadband Dielectric Spectrometer to characterize the electrical properties of a film metalized with thermally evaporated aluminum. The frequency $\nu$ window in the measurement was between 0.3Hz and 1MHz. Since polyetherimide is prone to absorb water \cite{Fontanella2007}, the sample was pre-heated under nitrogen flow to 200$\degC$C to remove water; this temperature was close to the reported glass transition temperature in the product sheet ($T_g=215$ by ASTM D3418). The temperature during the measurements were between -110$\degC$C and 270$\degC$C. 

\section{Electrical properties of polyetherimide}\label{sec:electr-prop-poly}

\subsection{Dielectric properties}

\begin{figure}[tp]
   \centering
   \includegraphics[width=.8\linewidth]{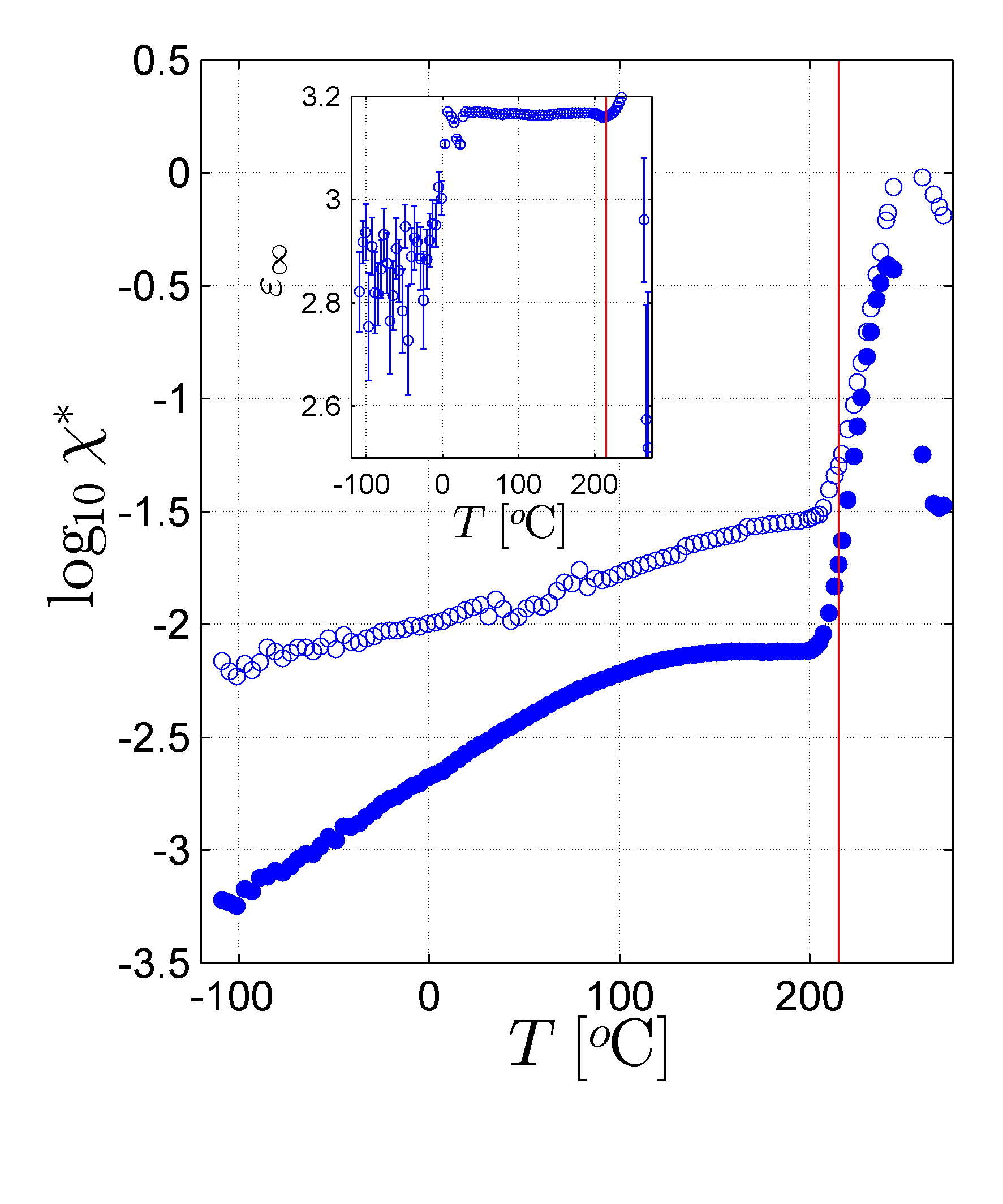}
   \caption{Complex dielectric susceptibility of polyetherimide as a function of temperature at 1kHz. The susceptibility is calculated using Eqs.~(\ref{eq:chi}) and (\ref{eq:drt}) with the analysis of the distribution of relaxation times approach. The open ($\circ$)and filled ($\bullet$) symbols represent the real and imaginary parts of the susceptibility. The inset shows the permittivity at high frequencies as a function of temperature calculated with the distribution of relaxation times approach. The solid vertical lines show the position of the glass transition $T_g$ temperature.}
   \label{fig:epsilonVSepsilon}
\end{figure}

The complex dielectric susceptibility $\chi^\ast$ of polyetherimide $5\mu$m thick film is shown for the data collected at 1kHz in Fig.~\ref{fig:epsilonVSepsilon} . The thin film thickness and the sensitivity of the capacitance measurement to the thickness the relative permittivity value of the polyetherimide is just over 3, which could be explained with the $\pm 10\%$ error in the thickness measurement for the film. The dielectric behavior of the film starts to change abruptly close to the glass transition temperature as expected. Isochronal measurement at 1kHz applied electric field clearly indicate the different relaxation in polyetherimide; one at low temperatures peaking around 120$\degC$C and one in at high temperatures around 230$\degC$C, see Fig.~\ref{fig:epsilonVSepsilon}. 

The complex susceptibility $\chi^*$ at 1kHz is shown in Fig.~\ref{fig:epsilonVSepsilon}. The susceptibility was calculated with the representation described in Eqs.~(\ref{eq:chi}) and (\ref{eq:drt}). The inset in the figure shows the high frequency value of the permittivity, which is not changing between 0$\degC$C and 200$\degC$C. Since the losses below 0$\degC$C are low the large oscillations in the relative permittivity at high frequencies were related to the numerics.

The polyetherimide is a low loss material below 200$^o$. The evolution of the segmental motion is clear, where the losses $\chi''$ increase starting from the lowest temperature measured (-110$\degC$C) to $200^o$C. The cooperative motion or cold liquid phase start to rearrange the polyetherimide molecules. The observation shows that the polyetherimide is a hot liquid with molecules easily change their confirmation over 255$\degC$C, where the real part of the susceptibility is decreasing and the imaginary part is becoming low again; few frictional losses; no polarizability over 255$\degC$C.

The low temperature process is the change in the segmental motion with temperature (assigned as the $\beta$-relaxation \cite{GlassTrans}) and is a broad relaxation. The $\alpha$-peak related to the cooperative motion of the polarizing units, which occur after the glass temperature, is much sharper in temperature axis and indicates the glass transition, softening of the polymer. The data presented by \citet{Belana1998Part1}  on the physical properties of polyetherimide had similar findings from the mechanical and dielectric studies for the $\beta$- and $\alpha$-relaxations but at lower temperatures because they have presented their data at $10Hz$ and $3Hz$ for electrical and mechanical measurement, respectively. The data presented by \citet{Diaz1998Part2} on the other hand presented dielectric loss data at different frequencies with labeled relaxations. The main difference between the current polyetherimide and the ones presented by \citet{Belana1998Part1} and \citet{Diaz1998Part2} is the $\gamma$-relaxation, which is not present in the studied material. This observation indicates that the tested current film did not have any side-groups, which was proposed to contribute to the dielectric relaxation at low temperatures\cite{Diaz1998Part2}. 

The space charge relaxation (denoted by $\rho$) mentioned by \citet{Belana1998Part1} could not be observed at 1kHz, however, our data show the presence of this relaxation at low frequency isochronal data, which was not shown here. One should mention that the $\rho$-relaxation is related to the blocking electrodes and is an indication of free ions; therefore related to the $\alpha$-relaxation. Observe that onset position of the increase in the dielectric losses ($\chi''$) and the increase in the relative permittivity at high frequencies $\varepsilon_\infty$ are just below the glass transition reported in the literature \cite{Diaz1998Part2,Belana1998}; the solid line in Fig.~\ref{fig:epsilonVSepsilon} show the reported glass transition $T_g=218-220^o$C.  The complete melting of the polymer is visible over 240$\degC$C, where the real and imaginary parts of the susceptibility are decreased quickly.  

\subsection{Relaxation map}

\begin{figure}[tp]
   \centering
   \includegraphics[width=.8\linewidth]{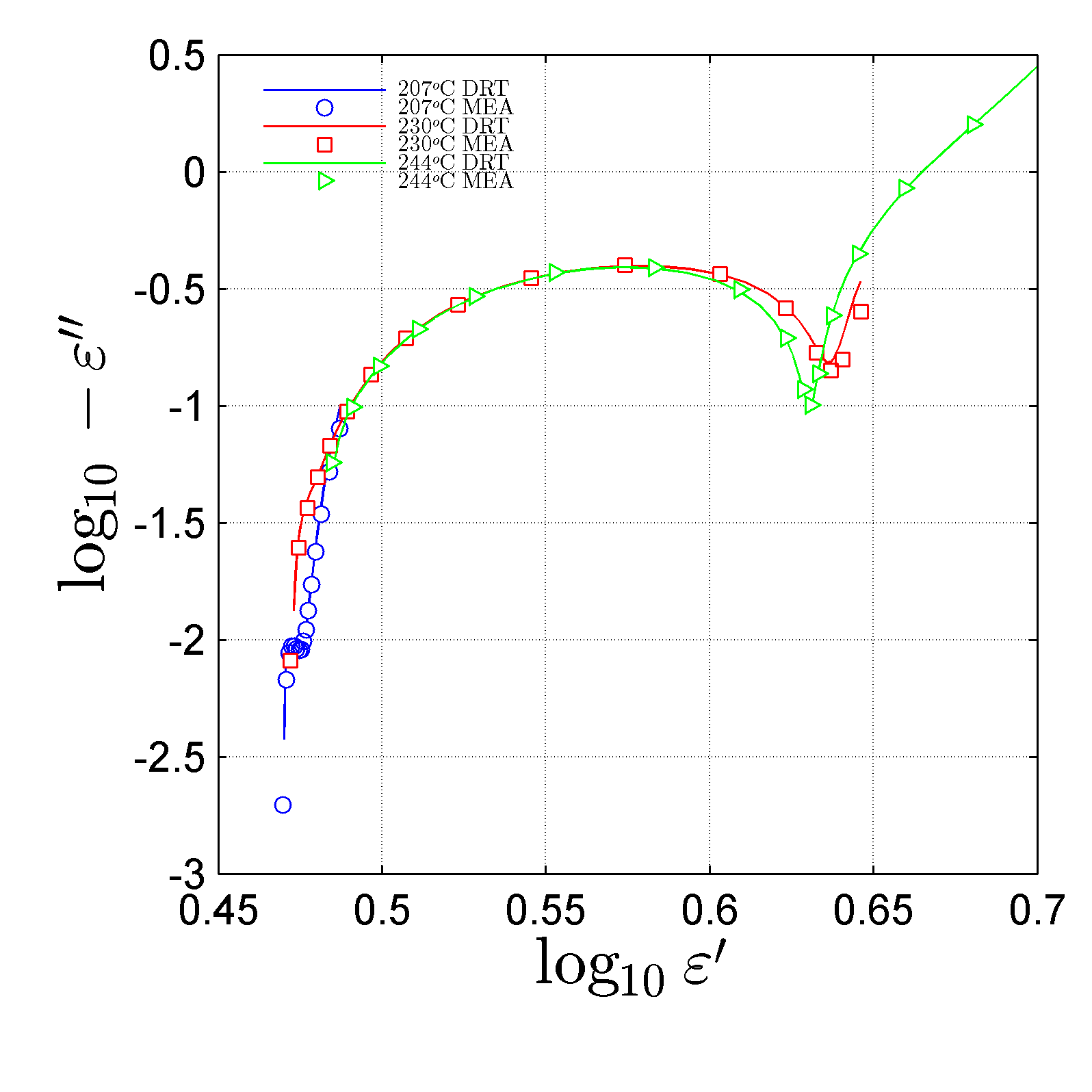}
   \caption{Argand plot of the complex dielectric permittivity at indicated temperatures below and above the glass transition. The symbols open circles ($\circ$), open squares ($\Box$) and open triangles ($\rhd$) indicate temperatures 207$\degC$C, 230$\degC$C and 244$\degC$C, respectively.}
   \label{fig:chiVStemp}
\end{figure}

\begin{figure}[tp]
   \centering
   \includegraphics[width=.8\linewidth]{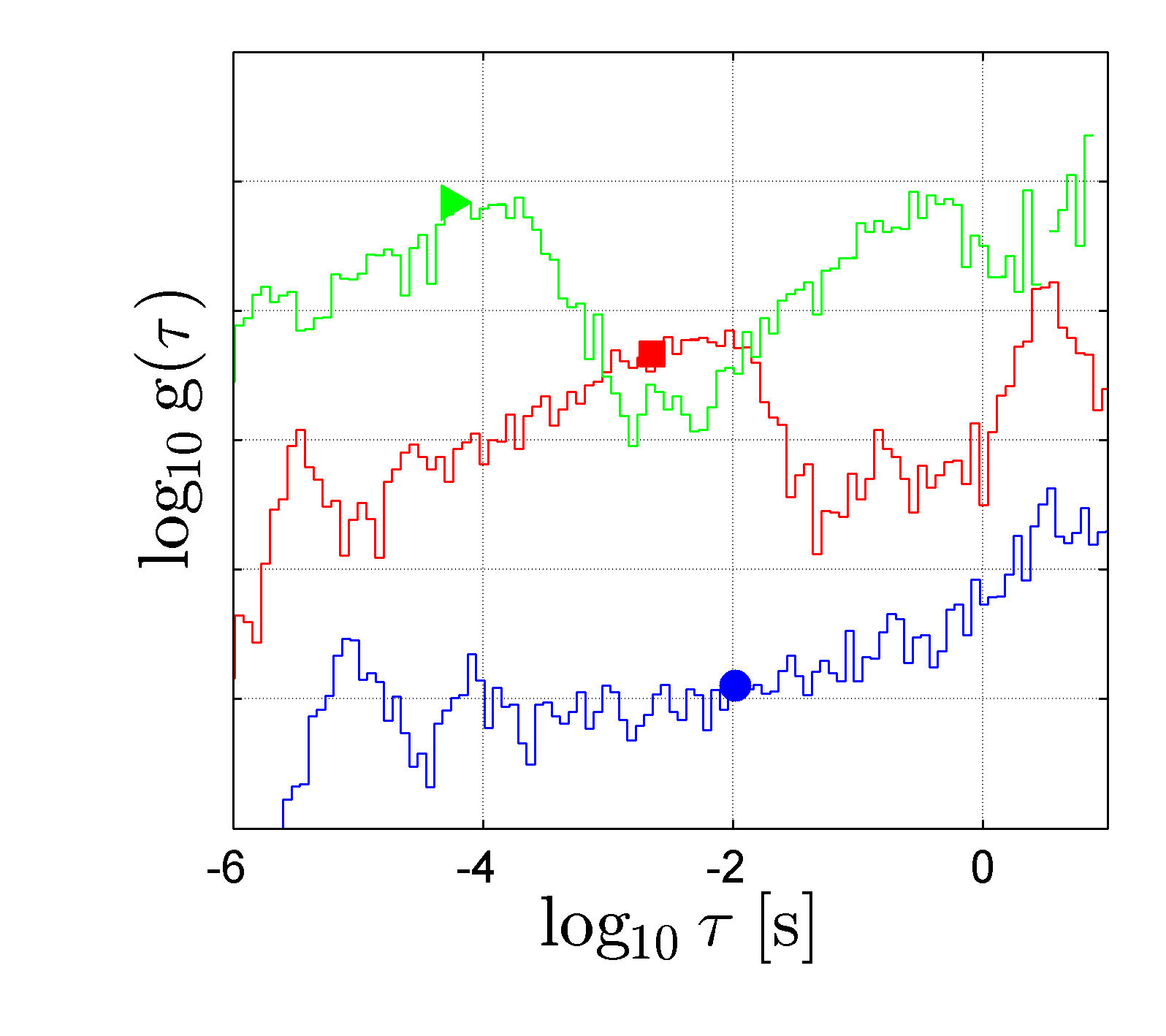}
   \caption{Distribution of relaxation times resolved for three temperatures as indicated in the plot. The distributions are shifted to guide the eyes. The data from bottom to top show the temperatures 207$\degC$C, 230$\degC$C and 244$\degC$C, respectively. The estimated expected relaxation times from Eq.~(\ref{eq:mostexptau}) are shown with filled symbols for the three temperatures.}
   \label{fig:dist_temp}
\end{figure}

The Argand plot of complex permittivity is shown in Fig.~\ref{fig:chiVStemp}. The relaxations just below and over $T_g$ are shown due to clear change in the dielectric response. The $\beta$-relaxation is starting to merge with the $\alpha$-relaxation. The resolved relaxation times are shown in semi-circle is due to the $\alpha$-relaxation. The data used to generate the responses were calculated using the distributions in Fig.~\ref{fig:dist_temp}. The peak positions for the distributions are estimated using Eq.~(\ref{eq:mostexptau}). Observe that the relaxation at 207$\degC$C (labeled with $\bullet$ in the figure) has no clear peak in the considered experimental window. However using the expression we were able to assign a most expected relaxation to it. The other temperature show relaxation peaks corresponding to $\alpha$- and $\beta$-relaxations.

The dielectric response of polyetherimide at three different temperatures are shown in Fig.~\ref{fig:chiVStemp}. The open symbols show the experimental values and the solid lines are the data obtained from the distribution of relaxations times approach. Below glass transition no semi-arcs are observed, however with the formation of cold liquid after the glass transition the cooperative motion peak was visible. The conductivity was also getting into the picture as the temperature was increased further due to increased mobility of ions and molecules.

The distribution of relaxation times for these temperatures are shown in Fig~\ref{fig:dist_temp}. The distribution at 207$\degC$C is nearly flat with some indication of slow relaxation moving into the measurement window. Clear peaks for the cooperative motion are observed at 230$\degC$C and 244$\degC$C. The filled symbols for each temperature indicates the most expected relaxation time $\bar{\tau}$ from Eq.~(\ref{eq:mostexptau}), which was used to build the relaxation map.

\begin{figure}[tp]
   \centering
   \includegraphics[width=.8\linewidth]{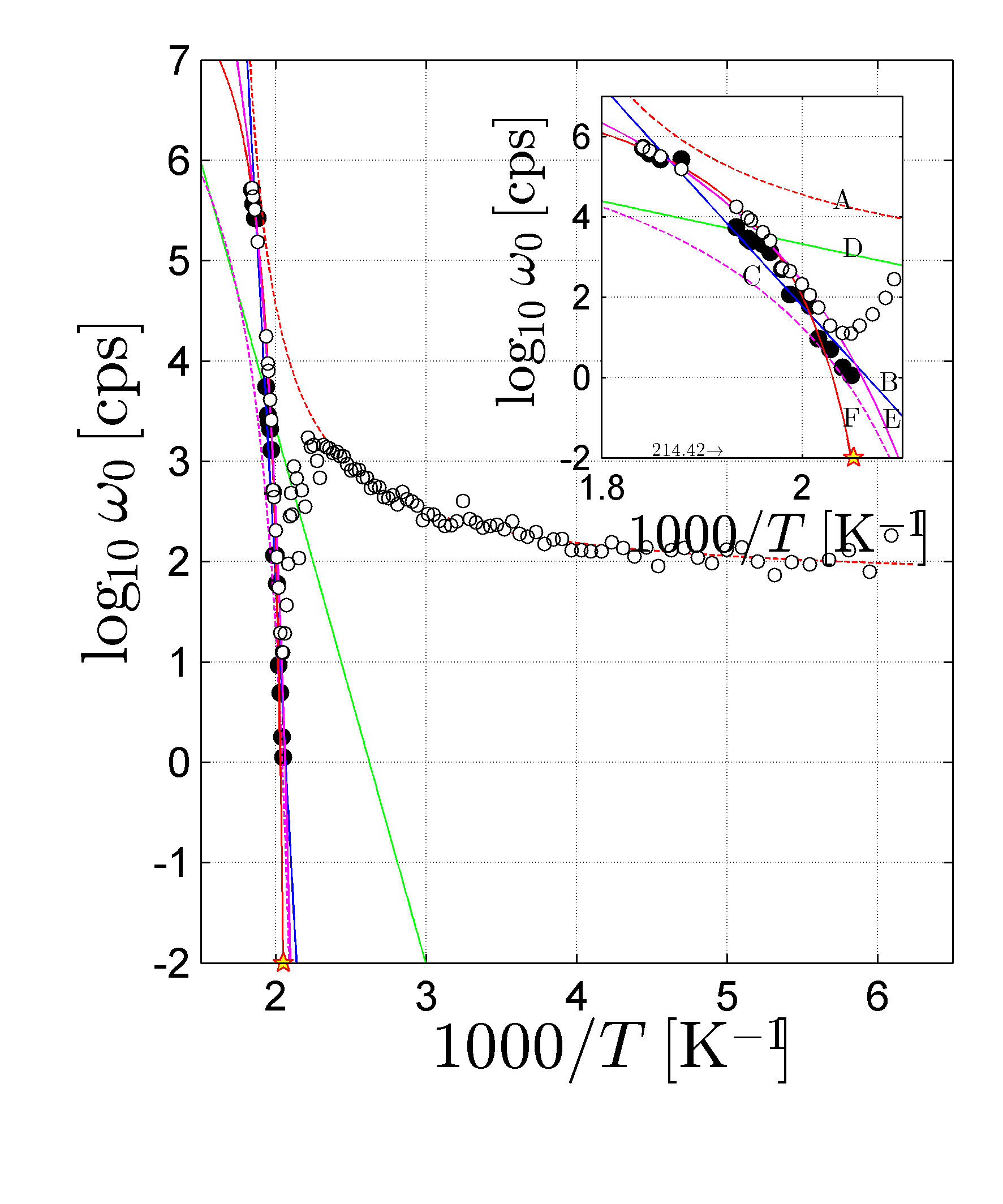} 
   \caption{Relaxation map for polyetherimide. The open symbols are the data from the expected relaxation times $\bar{\tau}$ and the filled symbols are taken for the maximum of the $\alpha$ relaxation. Curves `F' and 'A' are the VTF expressions for the $\alpha$ and $\beta$ relaxations, respectively. The data obtained from the peak positions were fitted with Arrhenius expression (curve 'B'), which is not as a good-fit as the VTF. Curve `D' is the data from \citet{Fontanella2007}. Curves `C' and 'F' are the WLF expressions, with constant $C_1$, $C_2$ and $T_0=T_g$ and optimized  $C_1$, $C_2$ and $T_0$ values, respectively.}
   \label{fig:relaxmap}
\end{figure}

Keeping track of the most expected relaxation times $\bar{\tau}$ at different temperature, we were  able to compile the relaxation map as shown in Fig. \ref{fig:relaxmap}. The filled and unfilled symbols used to differentiate between $\max [\rm{g}(tau)]$ and results from Eq.~(\ref{eq:mostexptau}) on the temperature region over $T_g$. Only the $\alpha$ relaxation yields maximum relaxation times at temperatures higher than the glass transition temperature $T_g$. The data from \citet{Fontanella2007} is also shown in the figure with solid lines (curve `D'). The literature data was not able to represent the relaxations resolved in our analysis, which might be due to slight differences in the raw materials and temperature range; we were able to perform measurements up to 270$\degC$C.  In Fig.~\ref{fig:relaxmap} the inset illustrates the cross-over region and the $\alpha$-relaxation to indicate differences in the expressions used in the data analysis.  No care was given here to separate $\alpha$ and $\beta$ relaxations as in previous studies \cite{Fontanella2007,mudarra:4807,Mudarra1999,Belana1998,suh:6333,378912}. 

Several different models were adopted to model the relaxation map. Arrhenius model has been applied to model the $\beta$ relaxation of many relaxing systems. In our analysis since we do not have a clear peak that moves on a straight line in the inverse temperature logarithm of relaxation rate representation for this relaxation in the distribution of relaxation times function $\rm{g}(\tau)$, $\beta$-relaxation was not considered with Arrhenius expression,
\begin{eqnarray}
  \label{eq:arrhenius}
  \omega_0=A\exp[-W_E(k_bT)^{-1}],
\end{eqnarray}
where, $A$ is the pre-exponential term usually around $10^{14}$s; $W_E$ is the activation energy in $e$V units; $k_b$ is the Boltzmann constant $k_b=8.6132\times 10^{-5}e$V; $T$ is the temperature in Kelvin. However, we have adopted Vogel-Tammann-Fulcher (VTF) equation\cite{Vogel,Tammann,Fulcher}, due to the bending of the curve to upward in the relaxation map. 
\begin{eqnarray}
  \label{eq:vtf}
  \omega_0=B\exp[D/(T-T')],
\end{eqnarray}
where, $B$ is the pre-exponential; $D$ is a fitting parameter in temperature units; $T'$ is the critical temperature $T>T'$. The segmental motion ($\beta$-relaxation) was modeled with $B=2.24\times10^{13}$ and $D=-885.96$ and $T'=612.35$ with negative temperature difference in the dominator. The VTF expression can represent the data to temperatures close to the merging region of the two relaxations; it is labeled with `A' in the inset of Fig.~\ref{fig:relaxmap}. After $\alpha$ and $\beta$ relaxations merge, it is interesting to see that the continuation of the fitted curve is also able to represent the $a$-relaxation, which is defined by \citet{GlassTrans} as the relaxation process at high temperatures after the two relaxation merged. Our modeling indicate that the segmental motion becomes faster as we approach to glass transition. This observation was expected since the thermal energy provided would not only provide kinetic energy to relaxing units, it also changes the landscape and might create free sites for the relaxing units to hop. It is hard to study thin samples at temperatures higher than the glass transition temperature due to phase change and melting of the polyetherimide, however, it would be valuable to record dielectric data at high temperature to attempt to follow the relaxations on this region $T>220^o$C in future studies. 

The $\alpha$ relaxation was modeled with four expressions; one Arrhenius, one VTF and two William-Landel-Ferry \cite{WLF} (WLF) models. The WLF model was adopted as proposed by the authors \cite{WLF} and also modified with free parameters. The WLF expression is as follows,
\begin{eqnarray}
  \label{eq:WLF}
  \ln\omega_0=C_1(T-T_0)[C_2+(T-T_0)],
\end{eqnarray}
where, $C_1$ and $C_2$ are fit parameters and $T_0$ is a reference temperature. The parameters $C_1$ and $C_2$ were given as 17.44 and 51.6, respectively\cite{WLF,GeddeBook}, if the reference temperature is taken to be the glass transition $T_g$ temperature (which is 215$\degC$C). This curve is illustrated in Fig.~\ref{fig:relaxmap} with label 'C'. It was not enough to express the relaxation behavior of polyetherimide over glass transition. 

When we took all the parameters free and applied a curve-fitting procedure, the parameters obtained were as follows $C_1=25.99$, $C_2=54.87$ and $T_0=485.51$, this curve is shown with E in Fig.~\ref{fig:relaxmap} inset. The glass transition temperature $T_g$ obtained from the WLF analysis is lower than the $T_g$ estimated with VTF, which is shown with `E' in the inset of Fig.~\ref{fig:relaxmap}. The VTF parameters obtained from the analysis were  $B=2.156\times10^8$, $D=-446.86$, and $T'=469.32$. Using the interpolation value at $\log_{10}\omega_0=-2$ for the $T_g$ estimate, the glass transition $T_g$ temperature of film polyetherimide is 214.57$\degC$C, which is close to the values published in the literature\cite{Fontanella2007,Diaz1998Part2,Belana1998Part1}. The data from \citet{Fontanella2007} for the $\alpha$-relaxation is plotted in Fig.~\ref{fig:relaxmap} and labeled as curve `D'. Their Arrhenius model was not representing our data. Note that  the volume recovery measurements performed on polyetherimide has reported a glass transition temperature $T_g$ of around 207.5$\degC$C.

\begin{figure}[tp]
   \centering
   \includegraphics[width=.8\linewidth]{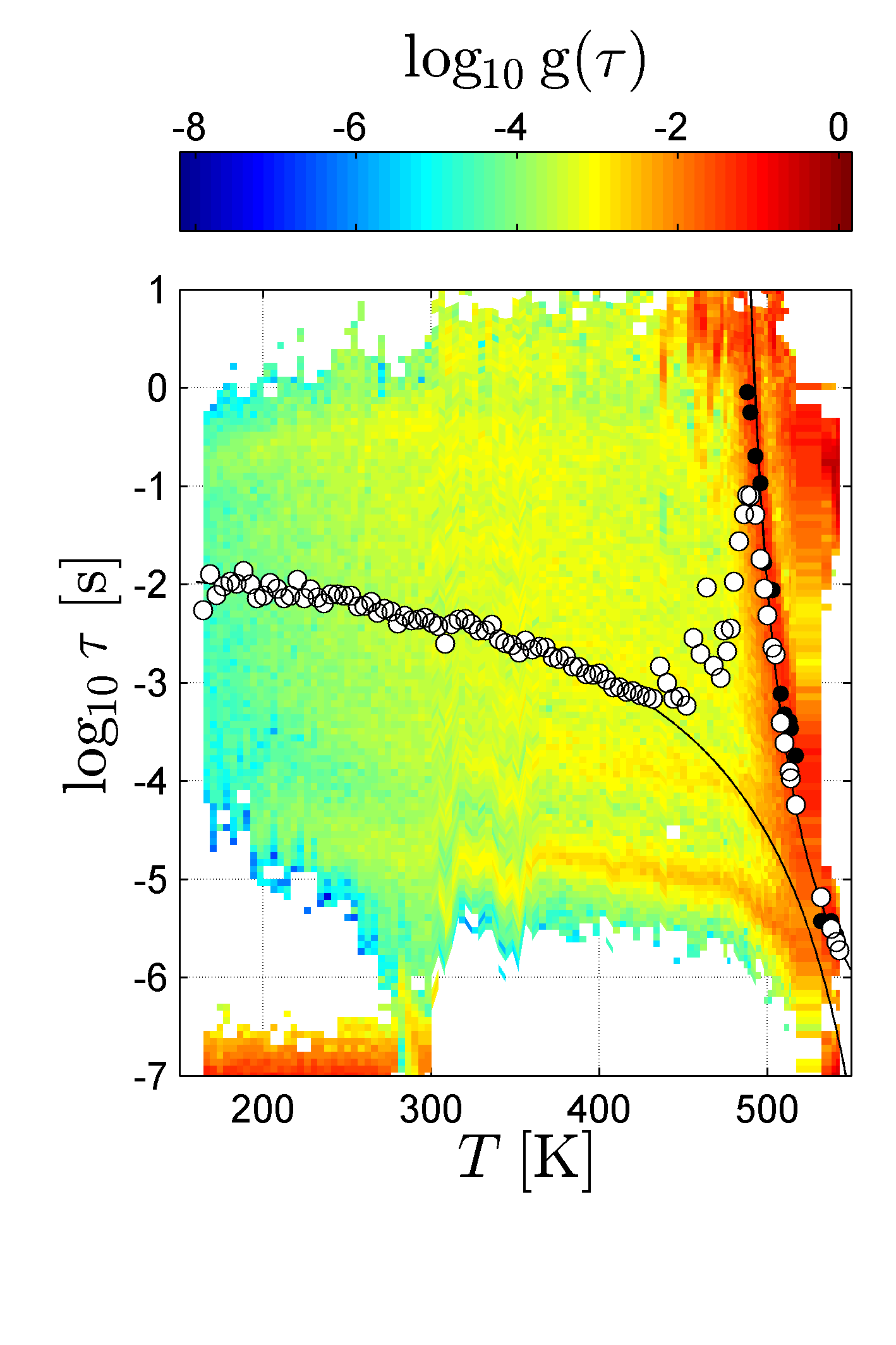}
   \caption{Relaxation map resolved using the distribution of relaxation times approach. The color code for the distribution of relaxation times map is shown on the top of the graph. The estimated glass transition temperature is 214.57$\degC$C from the VTF model, Eq.~(\ref{eq:vtf}).}
   \label{fig:relaxationmap}
\end{figure}

The full relaxation map estimated by the distribution of relaxation times and and the single point expected relaxation time estimates are shown in Fig.~\ref{fig:relaxationmap}. The relaxation times for the $\alpha$ process is much sharper than the $\beta$ one, which is nearly flat below the glass transition $T_g$. The relaxations resolved for $\beta$ resembles `fingers' coming out the $\alpha$-relaxation. Below room temperature ($<300$K), there are no peaks in the map. Influence of experimental error on the numerical error in the adopted inversion procedure has been estimated numerous times and shown that the inversion method not sensitive to the experimental data.  The one explanation to the `fingers' could be that these oscillating sort of behavior is due to the flat distribution as proposed by \citet{Frohlich}, who employed an arbitrary box distribution. Such a distribution was tested in the literature previously with the current analysis method \cite{Tuncer2004a}. The results have shown in that one might get oscillations of this sort. However, in the polyetherimide case the oscillations were more significant.
 
\subsection{Fragility of polyetherimide}
The fragility of glass-forming molecular systems was studied by \citet{AngellNgai,Angell1985}. Non-Arrhenius temperature dependence of the relaxation is defined as the dynamic fragility. It refers to deviations from Arrhenius temperature and expressed with VTF \cite{Vogel,Tammann,Fulcher} (Eq.~(\ref{eq:vtf})) or WLF \cite{WLF} (Eq.~(\ref{eq:WLF})) behavior. The fragility $m$  is estimated as,
\begin{eqnarray}
  m&=&{\rm d}(-\log_{10}{\omega_0})[{\rm d}(T_g/T))^{-1}|_{T\rightarrow T_g}  \label{eq:frag}\\ 
  &\equiv& DT_g(T_g-T')^{-2}\log_{10}e|_{T\rightarrow T_g}\quad \text{for Eq.~(\ref{eq:vtf})} \label{eq:fragVTF}\\
  &\equiv& C_1T_g(C_2)^{-1}|_{T\rightarrow T_g}\quad \text{for Eq.~(\ref{eq:WLF})} \label{eq:fragWLF}
\end{eqnarray}
The fragility index $m$ indicates how rapidly the molecular system changes as it approaches to the glass transition temperature from high temperature side of $T_g$; similarly just after passing $T_g$ from low temperature side, the molecular system would re-arrange (re-configure) itself, which is abruptly in the polyetherimide. For high $T_g$ thermoplastic polymers this change would yield to melting or even decomposition after glass transition. 

The polyetherimide has a high fragility $m$ around 284 and 230 that are estimated from Eq.~(\ref{eq:fragVTF}) and Eq.~(\ref{eq:fragWLF}). These values are much higher to those reported for other polymers in the literature \cite{HartmanBDSBook,Sokolov2007,GlassTrans,Qin20062977}. The difference in our estimations using VTF and WLF models should be understood in the view of how well these expressions describe the experimental data; as shown previously VTF was a better approach to the data. {\em Notice that the idea of using different curve fitting models is to better describe the data and avoid any numerical derivation, which is not straight forward. One can use polynomial approach as well however it has not been applied in the dielectric relaxation studies probably due to its non-physical basis.} 

As mentioned by \citet{Sokolov2007} high temperature polymers would have high fragility, which is mainly due to the high glass transition temperature $T_g$ and their rigidity. Using the proposed expression for the fragility index by \citet{Qin20062977} $$m\sim0.28(\pm0.067)T_g+9(\pm20),$$ we obtain $m\sim200(\pm32)$, which is lower than the value we have estimated. However, it yields a high fragility index.  The fragility index of polyetherimide was reported\cite{Simon1996_PEI_Fragility} as 214 for polyetherimide estimated with WLF model Eq.~(\ref{eq:fragWLF})(also listed in the table provide by \citet{Qin20062977}).
	
\subsection{Conductivity of polyetherimide}

\begin{figure}[tp]
   \centering
   \includegraphics[width=.8\linewidth]{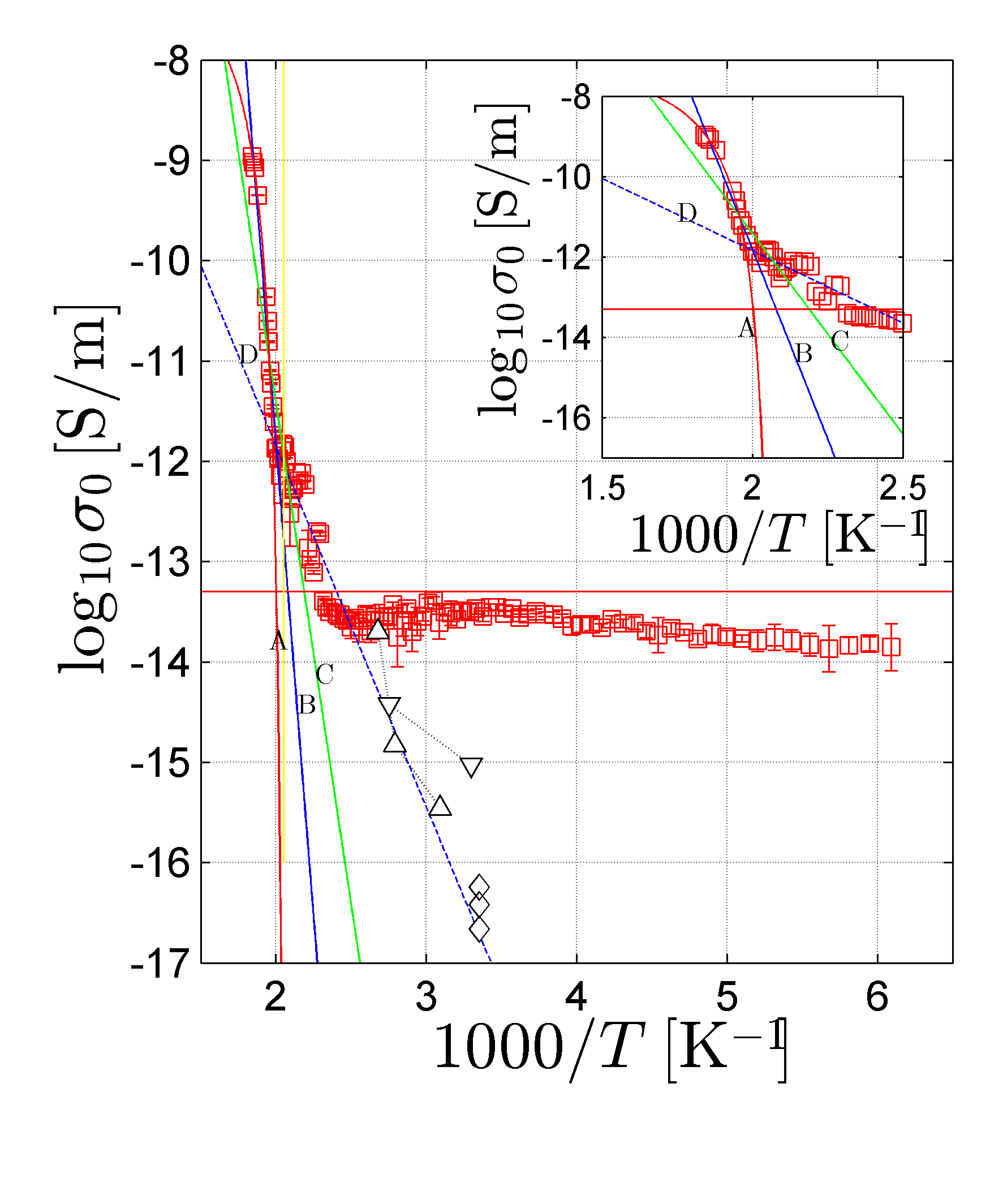}
   \caption{Ohmic conductivity of polyetherimide as a function of inverse temperature. The inset illustrates the region below and above the glass transition $T_g$ for polyetherimide. The horizontal solid line indicates the accuracy in the numerical analysis due to the considered frequency window. The curves that are presented represent different models; `A' is the VTF model for conductivity over $T>T_g$; `B' is Arrhenius fit for the same region; `C' is Arrhenius fit from \citet{mudarra:4807}; `D' is the Arrhenius model for the conductivity just below glass transition temperature $T<T_g$. The symbols are taken from the literature; ($\bigtriangleup$) from \citet{Zebouchi1998}; ($\bigtriangledown$) from \citet{suh:6333}; ($\Diamond$) from \citet{Reed1992}. }
   \label{fig:sigma0}
\end{figure}

Conduction processes in polyetherimide have been reported in the literature by several different groups \cite{suh:6333,mudarra:4807,Zebouchi1998,Reed1992,Fontanella2007}. As mentioned previously the distribution of relaxation times approach inherently performs or checks the Kramer-Kronig relations\cite{KramerKK,KramerKK2,KronigKK}, therefore the ohmic (direct current) conductivity is estimated directly, and value does not involve any electrode polarization losses due to low frequency polarization processes. The conductive as a function of inverse temperature is shown in Fig.~\ref{fig:sigma0}. Due to the limitation imposed by the selection of low frequency limit in the measurements, the conductivity values  $\sigma_0<1.67\times10^{-14}$Sm$^{-1}$ or $\log_{10}\sigma_0<-13.8$ would not be possible to estimate with any numerical method using the current instrument; more sensitive measurement techniques are needed to resolve such low conductivities. This is clearly presented in Fig.~\ref{fig:sigma0} as a flat conductivity estimate below $1000/T=2.5$ or 130$\degC$C. Over 130$\degC$C an increase in the conductivity is observed. This increase in the conductivity changes after the glass transition temperature indicating the change in the mobility in charge carriers after $T_g$. A similar behavior was observed for $\alpha$-polyvinylidene fluoride\cite{TuncerJCP2005}.

We have adopted two equations to describe the conduction in polyetherimide; Arrhenius and VTF expressions in Eq.~(\ref{eq:arrhenius}) and Eq.~(\ref{eq:vtf}) by converting $\omega_0$'s to $\sigma_0$'s in the expression, respectively. Data from \citet{mudarra:4807} was included in the conductivity plot (curve `C'), which was generated with $A=3.6\times10^8$Sm$^{-1}$ and $W_E=1.98e$V in Eq.~(\ref{eq:arrhenius}); it was only applicable to a limited temperature range. While the other data on polyetherimide conductivity from the literature\cite{Zebouchi1998,Reed1992,suh:6333} were also analyzed with our results. 

The curve generated to represent the data for temperature just below glass transition (Curve `D') yields close conductivity values, when extrapolated to low temperatures, to those given by \citet{Zebouchi1998} and \citet{Reed1992}. The curve was generated with  $A= 2.31\times10^{-5}$ and $W_E=0.71$. While the data from \citet{suh:6333} had one data point matching our results. Observe that the data of \citet{Reed1992} is 3 decades lower than that could be resolved with the measurements with the lowest frequency chosen (0.3Hz) and the numerical technique. To observe such low current values, we must have employed several $\mu$Hz frequency in our measurements. 

Since the fragility of polyetherimide is high, at high temperatures the behavior of conductivity could both be represented with VTF and Arrhenius, due to steep change. Both curves `A' and `B' were obtained from optimization with $B=1.50\times10^{-7}$, $D=315.8$ and $T'=478.11$, and $A=8.82\times10^{25}$Sm$^{-1}$ and $W_E=-3.74e$V, respectively. The VTF expression was a better fit due to its ability to model high temperature regions. The estimated VTF temperatures for the $\alpha$-relaxation and the conductivity ($T'$s) are close to each other indicating that the conduction in polyetherimide is ionic assisted by the cooperative motion, which determines the movement of the charge carriers. This observation was similar to $\alpha$-polyvinylidene fluoride\cite{TuncerJCP2005}.
 
\section{Discussion and conclusions}\label{sec:disc-concl}

An attempt to characterize dielectric response of thin polyetherimide films was presented here. A numerical method based on an numerical inversion algorithm was applied to resolve the distribution of relaxation times in 5$\mu$m thick polyetherimide film. The employed method, as shown several times previously  \cite{TuncerINTECH2011,TuncerMacdonald2007,Nanotech2006,TuncerPhilMag2006,TuncerJAP2006,TuncerJNCS,TuncerJCP2005,Tuncer2004a,Tuncer2002b,Tuncer2002elec,TuncerPhD,Tuncer2000b}, does not suffer from {\em a-priori} assumptions like in other non-linear fitting procedures \cite{NelderMead,LEVM}. It is therefore more straight forward for the user to analyze large data sets as the one presented here; broadband dielectric response of a polymer at different temperatures. However, post-processing of the data needs improvement for representation and storage. In conventional curve fitting methods, although the fitted curves are not excellent, the number of parameters stored are scarce. 

Notice that the nonparametric distribution of relaxation times approach brings in new insights to the dielectric relaxation as shown by \citet{TuncerPhilMag2006}, where the influence of coated hollow glass sphere particles on the dielectric response of paraffin wax were determined for electromagnetic shielding applications. The accuracy and strength of the numerical method has been discussed by   \citet{TuncerJAP2006} and \citet{TuncerMacdonald2007}, where the method employed here was benchmarked with a dielectric-community-excepted-free-software LEVM \cite{LEVM}. The nonparametric approach enable removing the observer (data analyst) from the data analysis. An analogy would be a spectroscopic measurement where one instrument scans over few energy spots to represent the whole spectrum using empirical function, while the another instrument scans the whole energy landscape to describe the system. Then the former is the conventional nonlinear fitting methods and the latter is the method presented here. Since the whole frequency region and the complex permittivity have been employed without a base function ({\em e.g.} Havriliak-Negami\cite{HNdist}, stretch exponential\cite{Kohl}, etc.), the data analyzed by two different analysts would be the same; no initial guess for the fit parameters are needed that would be different for the analysts.

Our analysis indicated that the tested film did not show any $\gamma$ relaxation as the ones reported\cite{Belana1998Part1,Diaz1998Part2}. Perhaps the current grade of polyetherimide was modified and did not have any side-groups, however other relaxations were present. Comparison of our relaxation map to the data from the literature revealed discrepancies; relaxation map from \citet{Fontanella2007} had the $\alpha$-relaxation in another location. This was attributed to the limited temperature range in their experiment.

The estimated direct-current conductivity of polyetherimide was compared to the data in the literature\cite{suh:6333,mudarra:4807,Zebouchi1998,Reed1992,Fontanella2007}. Good agreement was obtained for the data below the glass transition temperature $T_g$. Extrapolation of the curve that described data matched with those of \citet{Zebouchi1998,Reed1992}, which were measured with time domain methods.

 The results of the dielectric response have shown that polyetherimide has superb properties for passive electrical component applications. Its high glass transition temperature $T_g\sim214$ and low loss behavior over a wide range of temperatures (up to 200$\degC$C) would be benefited for high temperature applications, primarily capacitors for transformational power systems. Low conductivity again up to 200$\degC$ is important for direct-current applications and energy storage because of nearly negligible leakage through the material in those conditions. However, the conductivity brings in other challenge as charging and dissipation of induced charge in free-standing films together with high dielectric permittivity compared to other thermoplastics used in energy storage applications.


	

\begin{thebibliography}{74}
\providecommand{\natexlab}[1]{#1}
\providecommand{\url}[1]{\texttt{#1}}
\expandafter\ifx\csname urlstyle\endcsname\relax
  \providecommand{\doi}[1]{doi: #1}\else
  \providecommand{\doi}{doi: \begingroup \urlstyle{rm}\Url}\fi

\bibitem[Wirth and Heath(1974{\natexlab{a}})]{UltemPatent1}
J.~G. Wirth and D.~R. Heath.
\newblock Process for making polyetherimides.
\newblock Technical Report 3,787,364, USPO, 1974{\natexlab{a}}.

\bibitem[Wirth and Heath(1974{\natexlab{b}})]{UltemPatent2}
J.~G. Wirth and D.~R. Heath.
\newblock Process for making polyetherimides.
\newblock Technical Report 3,838,097, USPO, 1974{\natexlab{b}}.

\bibitem[Shugg(1986)]{Shugg1986}
W.~T. Shugg.
\newblock \emph{Handbook of electrical and electronic insulating materials}.
\newblock Van Norstrand Reinhold Company Inc., New York, 1986.

\bibitem[Belana et~al.(1998{\natexlab{a}})Belana, Cañadas, Diego, Mudarra,
  Díaz-Calleja, Friederichs, Jaïmes, and Sanchis]{Belana1998Part1}
J.~Belana, J.~C. Cañadas, J.~A. Diego, M.~Mudarra, R.~Díaz-Calleja,
  S.~Friederichs, C.~Jaïmes, and M.~J. Sanchis.
\newblock Comparative study of mechanical and electrical relaxations in
  poly(etherimide). part 1.
\newblock \emph{Polymer International}, 46\penalty0 (1):\penalty0 11--19,
  1998{\natexlab{a}}.

\bibitem[Diaz-Calleja et~al.(1998)Diaz-Calleja, Friederichs, Jaimes, Sanchis,
  Belana, Canadas, Diego, and Mudarra]{Diaz1998Part2}
R.~Diaz-Calleja, S.~Friederichs, C.~Jaimes, M.~J. Sanchis, J.~Belana, J.~C.
  Canadas, J.~A. Diego, and M.~Mudarra.
\newblock Comparative study of mechanical and electrical relaxations in
  poly(etherimide) part 2.
\newblock \emph{Polymer International}, 46:\penalty0 20--28, 1998.

\bibitem[Krause et~al.(1998)Krause, Yang, and Sessler]{Krause1998}
E.~Krause, G.~M. Yang, and G.~M. Sessler.
\newblock Charge dynamics and morphology of ultem 1000 and ultem 5000 pei grade
  films.
\newblock \emph{Polymer International}, 46:\penalty0 59--64, 1998.

\bibitem[Sarjeant et~al.(1998)Sarjeant, Zirnheld, and MacDougall]{Sarjeant1998}
W.~J. Sarjeant, J.~Zirnheld, and F.~W. MacDougall.
\newblock Capacitors.
\newblock \emph{IEEE Trans. Plasma Sci.}, 26\penalty0 (5):\penalty0 1368--1392,
  1998.

\bibitem[Reed and Cichanowski(1994)]{Reed1994}
C.~W. Reed and S.~W. Cichanowski.
\newblock The fundamentals of aging in hv polymeric-film capacitors.
\newblock \emph{IEEE Trans. Dielectr. Electr. Insulation}, 1\penalty0
  (5):\penalty0 904--922, 1994.

\bibitem[Rzad et~al.(1992)Rzad, Gasworth, Reed, and DeVre]{Reed1992}
S.J. Rzad, S.M. Gasworth, C.W. Reed, and M.W. DeVre.
\newblock Advanced materials for high energy density capacitors.
\newblock In \emph{Power Sources Symposium, 1992., IEEE 35th International},
  pages 358 --362, jun 1992.

\bibitem[Fontanella et~al.(2007)Fontanella, Bendler, Schuele, Edmondson, and
  Lomax]{Fontanella2007}
J.~J. Fontanella, J.~T. Bendler, D.~E. Schuele, C.~A> Edmondson, and J.~F.
  Lomax.
\newblock Effect of pressure on the water relaxation in glassy polyetherimide.
\newblock \emph{J. Non-Crystalline Solids}, 353:\penalty0 4528--4532, 2007.

\bibitem[Donth(2001)]{GlassTrans}
E.~Donth.
\newblock \emph{The Glass Transition: Relaxation Dynamics in Liquids and
  Disordered Materials}, volume~48 of \emph{Springer Series in Materials
  Science}.
\newblock Springer-Verlag, Berlin, 2001.

\bibitem[Lunkenheimer and Loidl(2003)]{LunkenheimerBDSBook}
P.~Lunkenheimer and A.~Loidl.
\newblock Glass dynamics beyond the $\alpha$-relaxation.
\newblock In F.~Kremer and A.~Sch{\"o}nhals, editors, \emph{Broadband
  Dielectric Spectroscopy}, chapter~14, pages 131--165. Springer-Verlag,
  Berlin, 2003.

\bibitem[Kremer and Sch{\"o}nhals(2003)]{KremerBook}
F.~Kremer and A.~Sch{\"o}nhals, editors.
\newblock \emph{Broadband Dielectric Spectroscopy}.
\newblock Springer-Verlag, Berlin, 2003.

\bibitem[Barsoukov and Macdonald(2005)]{MacdonaldImp2}
E.~Barsoukov and J.~R. Macdonald, editors.
\newblock \emph{Impedance Spectroscopy: Theory, Experiment and Applications}.
\newblock John Wiley \& Sons, New York, 2nd edition, 2005.

\bibitem[Blaise and Le~Gressus(1990)]{Blaise1990}
G.~Blaise and C.~Le~Gressus.
\newblock Charging phenomena, dielectric relaxation processes and breakdown of
  oxides.
\newblock In \emph{Electrical Insulation and Dielectric Phenomena, 1990. Annual
  Report., Conference on}, pages 231 --236, oct 1990.

\bibitem[Jonassen(2002)]{NielsElectrostatics}
Niels Jonassen.
\newblock \emph{Electrostatics}.
\newblock The Kluwer International Series in Engineering and Computer Science.
  Kluwer Academic Publishers, Norwell, Massachusetts USA, 2002.

\bibitem[Fr{\"o}hlich(1958)]{Frohlich}
H.~Fr{\"o}hlich.
\newblock \emph{Theory of Dielectrics; Dielectric constant and dielectric
  loss}.
\newblock Oxford Science Publications, Oxford, second edition, 1958.

\bibitem[Yager(1936)]{Yager1936}
W.~A. Yager.
\newblock The distribution of relaxation times in typical dielectrics.
\newblock \emph{Physics}, 7:\penalty0 434--450, 1936.

\bibitem[Kauzmann(1942)]{Kauzmann}
W.~Kauzmann.
\newblock Dielectric relaxation as a chemical rate process.
\newblock \emph{Rev. Mod. Phys.}, 14:\penalty0 12--44, 1942.

\bibitem[Macdonald and Barlow(1963)]{DRTMacdonals1963}
J.~R. Macdonald and C.~A. Barlow.
\newblock Relaxation, retardation, and superposition.
\newblock \emph{Rev. Mod. Phys.}, 35:\penalty0 940--946, Oct 1963.
\newblock \doi{10.1103/RevModPhys.35.940}.
\newblock URL \url{http://link.aps.org/doi/10.1103/RevModPhys.35.940}.

\bibitem[Jonscher(1983)]{Jonscher1983}
A.~K. Jonscher.
\newblock \emph{Dielectric Relaxation in Solids}.
\newblock London: Chelsea Dielectric, London, 1983.

\bibitem[McCrum et~al.(1967)McCrum, Read, and Williams]{McCrumDRT}
N.~G. McCrum, B.~E. Read, and G.~Williams.
\newblock \emph{Anelastic and Dielectric Effects in Polymeric Solids}.
\newblock John Wiley \& Sons Ltd., London, dover edition, 1967.

\bibitem[Kliem et~al.(1988)Kliem, Fuhrmann, and Arlt]{Kliem88}
H.~Kliem, P.~Fuhrmann, and G.~Arlt.
\newblock A numerical method for the determination of first-order kinetics
  relaxation time spectra.
\newblock \emph{IEEE Transactions on Electrical Insulation}, 23\penalty0
  (6):\penalty0 919--927, 1988.

\bibitem[Bella et~al.(1999)Bella, Laredo, and Grimau]{Grimau1999}
A.~Bella, E.~Laredo, and M.~Grimau.
\newblock Distribution of relaxation times from dielectric spectroscopy using
  monte carlo stimulated annealing: {A}pplication to $\alpha$-pvdf.
\newblock \emph{Phys. Rev. B}, 60\penalty0 (18):\penalty0 12764--12774, 1999.

\bibitem[Keiter and Rosenberg(1998)]{Keiter98}
H.~Keiter and M.~Rosenberg.
\newblock On the probability distributions of relaxation times in glasses.
\newblock \emph{European Journal of Physics B}, 5:\penalty0 599--603, 1998.

\bibitem[Dias(1996)]{Dias}
C.~J. Dias.
\newblock Determination of a distribution of relaxation frequencies based on
  experimental relaxation data.
\newblock \emph{Phys. Rev. B}, 53\penalty0 (21):\penalty0 14212--14222, 1996.

\bibitem[Havriliak and Negami(1966)]{HNdist}
S.~Havriliak and S.~Negami.
\newblock A complex plane analysis of {$\alpha$}-dispersion in some polymer
  systems.
\newblock \emph{J. Polym. Sci.: Part C}, 14:\penalty0 99--117, 1966.
\newblock The distribution ${\sf g}$ of depolarization factors is expressed as
  \begin{equation} \label{eq:havneg_dist} {\sf g}(s)=\frac{1}{\pi} \left |
  \frac{10^{s\alpha\beta}\sin(\beta\Theta)}{[10^{2s\alpha}+2 \cdot
  10^{s\alpha}\cos(\alpha\pi)+1]^{\beta/2}} \right | \end{equation} where,
  \begin{displaymath} \label{eq:havnew_theta} \Theta=\arctan \left [
  \frac{\displaystyle \sin(\alpha\pi)}{\displaystyle
  10^{s\alpha}+\cos(\alpha\pi)} \nonumber \right ] \end{displaymath} and
  $s=\log(z/{x})$ with ${x}$ being the most probable spectral parameter.

\bibitem[Arndt et~al.(1996)Arndt, Stannarius, Gorbatschow, and Kremer]{Arndt1}
M.~Arndt, R.~Stannarius, W.~Gorbatschow, and F.~Kremer.
\newblock Dielectric investigation of the dynamic glass transition in
  nanopores.
\newblock \emph{Phys. Rev. B}, 54\penalty0 (5):\penalty0 5377--5390, 1996.

\bibitem[Macdonald(2000{\natexlab{a}})]{Macdonald2000a}
J.~R. Macdonald.
\newblock Comparison of parametric and nonparametric methods for the analysis
  and inversion of immittance data: Critique of earlier work.
\newblock \emph{J. Comp. Phys.}, 157:\penalty0 280--301, 2000{\natexlab{a}}.

\bibitem[Macdonald(2000{\natexlab{b}})]{Macdonald2000b}
J.~R. Macdonald.
\newblock On relaxation-spectrum estimation for decades of data: accuracy and
  sampling-localization considerations.
\newblock \emph{Inv. Problems}, 16:\penalty0 1561--1583, 2000{\natexlab{b}}.

\bibitem[Macdonald(1995)]{macdonald:6241}
J.~R. Macdonald.
\newblock Exact and approximate nonlinear least-squares inversion of dielectric
  relaxation spectra.
\newblock \emph{J. Chem. Phys.}, 102\penalty0 (15), 1995.

\bibitem[Tuncer and Guba{\'n}ski(2001)]{Tuncer2000b}
E.~Tuncer and S.~M. Guba{\'n}ski.
\newblock On dielectric data analysis: {I}ntroduction of the {M}onte {C}arlo
  method to obtain distributions of relaxation times and a comparison with a
  functional approach.
\newblock \emph{IEEE Trans. Dielect. Elect. Insul.}, 8:\penalty0 310--320,
  2001.

\bibitem[Tuncer and Macdonald(2006)]{TuncerJAP2006}
E.~Tuncer and J.~R. Macdonald.
\newblock Comparison of methods for estimating continuous distributions of
  relaxation times.
\newblock \emph{J. Appl. Phys.}, 99:\penalty0 074106, 2006.

\bibitem[Macdonald and Tuncer(2007)]{TuncerMacdonald2007}
J.~R. Macdonald and E.~Tuncer.
\newblock Deconvolution of immittance data: Some old and new methods.
\newblock \emph{J. Elctroanal. Chem.}, 602:\penalty0 255--262, 2007.

\bibitem[Tuncer et~al.(2011)Tuncer, Belattar, Achour, and
  Brosseau]{TuncerINTECH2011}
E.~Tuncer, J.~Belattar, M.~E. Achour, and C.~Brosseau.
\newblock Broadband spectral analysis of non-debye dielectric relaxation in
  percolating heterostructures.
\newblock In Brahim Attaf, editor, \emph{Advances in Composite Materials for
  Medicine and Nanotechnology}, pages 1--12. InTech, 2011.
\newblock doi: 10.5772/14680.

\bibitem[Tuncer et~al.(2007)Tuncer, Sauers, James, Ellis, Paranthaman,
  Aytu{\u{g}}, Sathyamurthy, More, Li, and Goyal]{Nanotech2006}
E.~Tuncer, I.~Sauers, D.~R. James, A.~R. Ellis, M.~P. Paranthaman,
  T.~Aytu{\u{g}}, S.~Sathyamurthy, K.~L. More, J.~Li, and A.~Goyal.
\newblock Electrical properties of epoxy resin based nano-composites.
\newblock \emph{Nanotechnology}, 18:\penalty0 025703 (6pp), 2007.

\bibitem[Tuncer et~al.(2006)Tuncer, Bowler, Youngs, and
  Lymer]{TuncerPhilMag2006}
E.~Tuncer, N.~Bowler, I.~J. Youngs, and K.~P. Lymer.
\newblock Investigating low-frequency dielectric properties of a composite
  using the distribution of relaxation times technique.
\newblock \emph{Phil. Mag.}, 86:\penalty0 2359 -- 2369, 2006.

\bibitem[Tuncer et~al.(2005{\natexlab{a}})Tuncer, Wegener, and
  Gerhard-Multhaupt]{TuncerJNCS}
E.~Tuncer, M.~Wegener, and R.~Gerhard-Multhaupt.
\newblock Distribution of relaxation times in $\alpha$-phase polyvinylidene
  fluoride.
\newblock \emph{J. Non-Cryst. Solids}, 351\penalty0 (33-36):\penalty0
  2917--2921, 2005{\natexlab{a}}.

\bibitem[Tuncer et~al.(2005{\natexlab{b}})Tuncer, Wegener, Fr{\"{u}}bing, and
  Gerhard-Multhaupt]{TuncerJCP2005}
E.~Tuncer, M.~Wegener, P.~Fr{\"{u}}bing, and R.~Gerhard-Multhaupt.
\newblock Origin of temperature dependent conductivity in
  $\alpha$-polyvinylidene fluoride.
\newblock \emph{J. Chem. Phys.}, 122:\penalty0 084901, 2005{\natexlab{b}}.

\bibitem[Tuncer et~al.(2004)Tuncer, Furlani, and Mellander]{Tuncer2004a}
E.~Tuncer, M.~Furlani, and B.-E. Mellander.
\newblock Resolving distribution of relaxation times in poly(propylene glycol)
  on the crossover region.
\newblock \emph{J. Appl. Phys.}, 95\penalty0 (6):\penalty0 3131--3140, 2004.

\bibitem[Tuncer et~al.(2002)Tuncer, Nettelblad, and Guba{\'n}ski]{Tuncer2002b}
E.~Tuncer, B.~Nettelblad, and S.~M. Guba{\'n}ski.
\newblock Non-debye dielectric relaxation in binary dielectric mixtures
  (50-50): {R}andomness and regularity in mixture topology.
\newblock \emph{J. Appl. Phys.}, 92\penalty0 (8):\penalty0 4612--4624, 2002.

\bibitem[Tuncer et~al.(2001)Tuncer, Guba{\'n}ski, and
  Nettelblad]{Tuncer2002elec}
E.~Tuncer, S.~M. Guba{\'n}ski, and B.~Nettelblad.
\newblock Electrical properties of $4\times4$ binary dielectric mixtures.
\newblock \emph{Journal of Electrostatics}, 56\penalty0 (4):\penalty0 449--463,
  2001.

\bibitem[Tuncer(2001)]{TuncerPhD}
E.~Tuncer.
\newblock \emph{Dielectric relaxation in dielectric mixtures}.
\newblock PhD thesis, Chalmers University of Technology, Gothenburg, Sweden,
  2001.

\bibitem[Debye(1945)]{Debye1945}
P.~Debye.
\newblock \emph{Polar Molecules}.
\newblock Dover Publications, New York, 1945.

\bibitem[Macdonald(1953)]{MacdonaldDebye}
J.~R. Macdonald.
\newblock A new model for the debye dispersion equations.
\newblock \emph{Phys. Rev.}, 91:\penalty0 412--412, 1953.

\bibitem[Tuncer and Guba{\'n}ski(2002)]{Tuncer2001d}
E.~Tuncer and S.~M. Guba{\'n}ski.
\newblock On numerical simulations of composite dielectrics in thermally
  stimulated conditions.
\newblock \emph{Turkish Journal of Physics}, 26:\penalty0 1--33, 2002.

\bibitem[Tuncer(2005)]{TuncerSpectralPRB}
E.~Tuncer.
\newblock Extracting spectral density function of a binary composite without
  a-priori assumption.
\newblock \emph{Phys. Rev. B}, 71:\penalty0 012101, 2005.
\newblock ({\it Preprint} cond-mat/0403243).

\bibitem[Tuncer and Lang(2005)]{TuncerLang}
Enis Tuncer and Sidney~B. Lang.
\newblock Numerical extraction of distributions of space-charge and
  polarization from laser intensity modulation method.
\newblock \emph{Appl. Phys. Lett.}, 86:\penalty0 071107, 2005.
\newblock ({\em Preprint} cond-mat/0409183).

\bibitem[Lang and Tuncer(2008)]{TuncerLang2008}
Sidney Lang and Enis Tuncer.
\newblock Comparison of techniques for solving the laser intensity modulation
  method (limm) equation.
\newblock \emph{J. Electroceramics}, 21:\penalty0 827--830, 2008.

\bibitem[Tuncer(2012)]{TuncerTDEI2012}
E.~Tuncer.
\newblock Distribution of relaxation times: An inverse problem.
\newblock \emph{Dielectrics and Electrical Insulation, IEEE Transactions on},
  19\penalty0 (4):\penalty0 1221 --1225, august 2012.
\newblock ISSN 1070-9878.
\newblock \doi{10.1109/TDEI.2012.6259994}.

\bibitem[Tuncer(2011)]{TuncerISE2011}
E.~Tuncer.
\newblock Inverse problems in dielectrics.
\newblock In \emph{Electrets (ISE), 2011 14th International Symposium on},
  pages 77 --78, aug. 2011.
\newblock \doi{10.1109/ISE.2011.6084990}.

\bibitem[Belana et~al.(1998{\natexlab{b}})Belana, Canadas, Diego, Mudarra,
  Diaz, Friederichs, Jaimes, and Sanchis]{Belana1998}
J.~Belana, J.~C. Canadas, J.~A. Diego, M.~Mudarra, R.~Diaz, S.~Friederichs,
  C.~Jaimes, and M.~J. Sanchis.
\newblock Physical ageing studies in polyetherimide ultem 1000.
\newblock \emph{Polymer International}, 46:\penalty0 29--32,
  1998{\natexlab{b}}.

\bibitem[Mudarra et~al.(2000)Mudarra, Belana, {n}adas, Diego, Sellar\`{e}s,
  D\'{\i}az-Calleja, and Sanch\'{\i}s]{mudarra:4807}
M.~Mudarra, J.~Belana, J.~C.~Ca\ {n}adas, J.~A. Diego, J.~Sellar\`{e}s,
  R.~D\'{\i}az-Calleja, and M.~J. Sanch\'{\i}s.
\newblock Space charge relaxation in polyetherimides by the electric modulus
  formalism.
\newblock \emph{Journal of Applied Physics}, 88\penalty0 (8):\penalty0
  4807--4812, 2000.

\bibitem[Mudarra et~al.(1999)Mudarra, Belana, Diaz-Calleja, Canadas, Diego,
  Sellares, and Sanchis]{Mudarra1999}
M.~Mudarra, J.~Belana, R.~Diaz-Calleja, J.C. Canadas, J.A. Diego, J.~Sellares,
  and M.J. Sanchis.
\newblock Relaxation of space charge in polyetherimide by dynamic electrical
  analysis and thermally stimulated depolarization currents.
\newblock In \emph{Electrets, 1999. ISE 10. Proceedings. 10th International
  Symposium on}, pages 71 --74, 1999.

\bibitem[Suh et~al.(1996)Suh, Nam, and Lim]{suh:6333}
Kwang~S. Suh, Jin~Ho Nam, and Kee~Joe Lim.
\newblock Electrical conduction in polyetherimide.
\newblock \emph{Journal of Applied Physics}, 80\penalty0 (11):\penalty0
  6333--6335, 1996.

\bibitem[Suthar et~al.(1993)Suthar, Stokes, Khachen, Laghari, and
  Hammoud]{378912}
J.L. Suthar, A.~Stokes, W.~Khachen, J.R. Laghari, and A.~Hammoud.
\newblock Statistical analysis of multistress aging studies of polyetherimide.
\newblock In \emph{Electrical Insulation and Dielectric Phenomena, 1993. Annual
  Report., Conference on}, pages 568 --573, oct 1993.

\bibitem[Vogel(1921)]{Vogel}
H.~Vogel.
\newblock Das temperaturabhaengigkeitsgesetz der viskositaet von
  fluessigkeiten.
\newblock \emph{Physikalische Zeitschrift Leipzig}, 22:\penalty0 645, 1921.

\bibitem[Tammann and Hesse(1926)]{Tammann}
G.~Tammann and W.~Hesse.
\newblock Die abhängigkeit der viscosität von der temperatur bie
  unterkühlten flüssigkeiten.
\newblock \emph{Zeitschrift f{\"u}r anorganische und allgemeine Chemie},
  156\penalty0 (1):\penalty0 245, 1926.

\bibitem[Fulcher(1925)]{Fulcher}
G.~S. Fulcher.
\newblock Analysis of recent measurements of the viscosity of glasses.
\newblock \emph{Journal of American Ceramic Society}, 8:\penalty0 339, 1925.

\bibitem[Williams et~al.(1955)Williams, Landel, and Ferry]{WLF}
M.~L. Williams, R.~F. Landel, and J.~D. Ferry.
\newblock The temperature dependence of relaxation mechanisms in amorphous
  polymers and other glass-forming liquids1.
\newblock \emph{Journal of American Chemical Society}, 77\penalty0
  (14):\penalty0 3701–3707, 1955.

\bibitem[Gedde(1995)]{GeddeBook}
U.~Gedde.
\newblock \emph{Polymer Physics}.
\newblock Kluwer Academics Press Publisher, Dordrecht, the Netherlands, 1995.

\bibitem[Angell et~al.(2000)Angell, Ngai, Kenna, McMillan, and
  Martin]{AngellNgai}
A.~A. Angell, K.~L. Ngai, G.~B.~Mc Kenna, P.~F. McMillan, and S.~W. Martin.
\newblock Relaxation in glassforming liquids and amorphous solids.
\newblock \emph{J. Appl. Phys.}, 88\penalty0 (6):\penalty0 3113--3157, 2000.

\bibitem[Angell(1985)]{Angell1985}
C.~A. Angell.
\newblock Relaxations in complex systems.
\newblock \emph{J. Non-Cryst. Solids}, 73:\penalty0 1--14, 1985.

\bibitem[Hartman et~al.(2003)Hartman, Fukao, and Kremer]{HartmanBDSBook}
L.~Hartman, K.~Fukao, and F.~Kremer.
\newblock Molecular dynamics in thin polymer films.
\newblock In F.~Kremer and A.~Sch{\"o}nhals, editors, \emph{Broadband
  Dielectric Spectroscopy}, chapter~14, pages 131--165. Springer-Verlag,
  Berlin, 2003.

\bibitem[Sokolov et~al.(2007)Sokolov, Novikov, and Ding]{Sokolov2007}
A.~P. Sokolov, V.~N. Novikov, and Y.~Ding.
\newblock Why many polymers are so fragile.
\newblock \emph{J. Phys.: Condens. Matter}, 19:\penalty0 205116 (8pp), 2007.

\bibitem[Qin and McKenna(2006)]{Qin20062977}
Qian Qin and Gregory~B. McKenna.
\newblock Correlation between dynamic fragility and glass transition
  temperature for different classes of glass forming liquids.
\newblock \emph{Journal of Non-Crystalline Solids}, 352\penalty0
  (28–29):\penalty0 2977 -- 2985, 2006.

\bibitem[Simon et~al.(1997)Simon, Plazek, Sobieski, and
  McGregor]{Simon1996_PEI_Fragility}
Sindee~L. Simon, Donald~J. Plazek, J.~William Sobieski, and Eric~T. McGregor.
\newblock Physical aging of a polyetherimide: Volume recovery and its
  comparison to creep and enthalpy measurements.
\newblock \emph{Journal of Polymer Science Part B: Polymer Physics},
  35\penalty0 (6):\penalty0 929--936, 1997.

\bibitem[Zebouchi et~al.(1998)Zebouchi, Truong, Essolbi, Se-Ondoua, Malec,
  Vella, Malrieu, Toureille, Schué, and Jones]{Zebouchi1998}
N.~Zebouchi, V.~H. Truong, R.~Essolbi, M.~Se-Ondoua, D.~Malec, N.~Vella,
  S.~Malrieu, A.~Toureille, F.~Schué, and R.~G. Jones.
\newblock The electric breakdown behaviour of polyetherimide films.
\newblock \emph{Polymer International}, 46\penalty0 (1):\penalty0 54--58, 1998.

\bibitem[Kramers(1926)]{KramerKK}
M.~H.~A. Kramers.
\newblock \emph{Nature (London)}, 117:\penalty0 775, 1926.
\newblock Royal Danish Academy of Sciences and Letters Note on Kramers remarks.

\bibitem[Kramers(1927)]{KramerKK2}
M.~H.~A. Kramers.
\newblock La diffusion de la lumiere par les atomes.
\newblock \emph{Atti. Congr. Intern. Fisici, Como}, 2:\penalty0 545--557, 1927.

\bibitem[Kronig(1926)]{KronigKK}
R.~{de}~L. Kronig.
\newblock On the theory of the dispersion of x-rays.
\newblock \emph{J. Opt. Soc. Am.}, 12:\penalty0 547, 1926.

\bibitem[Nelder and Mead(1965)]{NelderMead}
J.~A. Nelder and R.~Mead.
\newblock A simplex method for function minimization.
\newblock \emph{Comput. J.}, 7:\penalty0 308--313, 1965.

\bibitem[Macdonald and L.~D.~Potter(1987)]{LEVM}
J.~R. Macdonald and {Jr} L.~D.~Potter.
\newblock A flexible procedure for analyzing impedance spectroscopy results:
  Description and illustrations.
\newblock \emph{Solid State Ionics}, 24\penalty0 (1):\penalty0 61--79, 1987.
\newblock the latest version of the LEVM fitting program, V. 8.0, may be
  obtained at no cost from http//www.physics.unc.edu/{$\sim$}macd/ where more
  details about the program appear. An extensive manual, source code, and
  executable code are included.

\bibitem[Kohlrausch(1854)]{Kohl}
R.~Kohlrausch.
\newblock Theorie des electrischen r\"uckstandes in der leidener flasche.
\newblock \emph{Annalen der Physik und der Physikalischen Chemie}, 91:\penalty0
  179--214, 1854.

\end{thebibliography}
	

\end{document}